\documentclass[aps,pra,reprint,twocolumn,superscriptaddress]{revtex4-1}

\usepackage{amsmath,graphicx,mathtools,amsfonts,epstopdf,hyperref}
\usepackage[caption=false]{subfig}

\newcommand{\ceil}[1]{\ensuremath{\lceil #1 \rceil}} 
\newcommand{\floor}[1]{\ensuremath{\lfloor #1 \rfloor}} 
\newcommand{\ket}[1]{\ensuremath{\left| #1 \right>}}
\newcommand{\bra}[1]{\ensuremath{\left< #1 \right|}}

\begin{document}

\title{\textbf{Entanglement witness via symmetric two-body correlations}}
\author{Ernest Y.-Z. Tan}
\affiliation{Centre for Quantum Technologies, National University of Singapore, 3 Science Drive 2, Singapore 117543, Singapore}
\affiliation{Department of Physics, National University of Singapore, 2 Science Drive 3, 117542 Singapore, Singapore}
\author{Dagomir Kaszlikowski}
\affiliation{Centre for Quantum Technologies, National University of Singapore, 3 Science Drive 2, Singapore 117543, Singapore}
\affiliation{Department of Physics, National University of Singapore, 2 Science Drive 3, 117542 Singapore, Singapore}
\author{L.C. Kwek}
\affiliation{Centre for Quantum Technologies, National University of Singapore,
3 Science Drive 2, Singapore 117543, Singapore}
\affiliation{Institute of Advanced Studies, Nanyang Technological University,
60 Nanyang View, Singapore 639673, Singapore}
\affiliation{National Institute of Education, Nanyang Technological University,
1 Nanyang Walk, Singapore 637616, Singapore}
\affiliation{MajuLab, CNRS-UNS-NUS-NTU International Joint Research Unit, UMI 3654, Singapore}

\begin{abstract}
We construct an entanglement witness for many-qubit systems, based on symmetric two-body correlations with two measurement settings. This witness is able to detect the entanglement of some Dicke states for any number of particles, and such detection exhibits some robustness against white noise and thermal noise under the Lipkin-Meshkov-Glick Hamiltonian. In addition, it detects the entanglement of spin-squeezed states, with a detection strength that approaches the maximal value for sufficiently large numbers of particles. As spin-squeezed states can be experimentally generated, the properties of the witness with respect to these states may be amenable to experimental investigation. Finally, we show that while the witness is unable to detect GHZ states, it is instead able to detect superpositions of Dicke states with GHZ states. 
\end{abstract}

\maketitle

\section{Introduction}

Quantum entanglement has been a useful resource for a number of interesting applications such as quantum cryptography~\cite{ekert91}, quantum teleportation~\cite{bennett93tele}, or even quantum error correction~\cite{bennett96qec}. In order to study the role of quantum entanglement in many of these applications,  a variety of measures have been introduced to detect or quantify entanglement, including concurrence, the Peres-Horodecki criterion, negativity, localizable entanglement, geometric measures and entanglement witnesses~\cite{hill97,peres96,horodecki96,popp05,terhal00}. In recent years, the study of quantum entanglement has been extended to the description and understanding of quantum many-body systems in condensed matter physics. Some techniques for investigating multipartite entanglement in such systems include the use of concurrence in symmetric states~\cite{vidal06} or spin-squeezing inequalities~\cite{sorenson01,korbicz06,toth07,toth09,lucke14}.

In a recent work, Tura~et~al.~\cite{tura14} derived a class of Bell inequalities~\cite{bell64} for multipartite systems based on symmetric one- and two-body correlations. Such considerations are useful since one- and two-body correlators are more easily accessible experimentally as compared to higher-order correlations. These Bell inequalities serve therefore as an important tool to demonstrate that some entangled states exhibit behaviour which cannot be described by local realistic models. However, it is known that not all entangled states are able to violate Bell inequalities directly~\cite{werner89}, and the precise details of the relationship between entanglement and Bell inequality violation are still being characterized~\cite{brunner14}. In this work, we use symmetric two-body correlations to provide an entanglement witness, and characterize some of its properties. In particular, it is able to detect the entanglement of some states which can be experimentally generated, including symmetric states such as spin-squeezed states and Dicke states~\cite{riedel10,gross10,raghavan01,vanderbruggen11,lucke14,milburn97,cirac98}. This casts some light on the relationship between spin-squeezing inequalities and the Bell inequalities proposed by Tura~et~al.~\cite{tura14}.

This paper is organized as follows: In section~\ref{MEW}, we introduce the multipartite entanglement witness with two-body correlations.  In section~\ref{GI}, we study the relation between the entanglement witness and the classical polytopes, and discuss differences between the cases of odd and even numbers of qubits. In section~\ref{Dicke}, we show that the ground states of the Lipkin-Meshkov-Glick (LMG) Hamiltonian are detected by the entanglement witness, and discuss the detection of thermal states associated with the LMG Hamiltonian. In section~\ref{Maximal}, we consider the states which are most strongly detected by the entanglement witness, and show that spin-squeezed states come close to fulfilling this property. In section~\ref{Superposed}, we consider detection of superpositions of Dicke states with GHZ states using the witness. Finally, we summarize our result in section \ref{Conclusion} with some brief remarks. 

\section{Multipartite entanglement witness}\label{MEW}

An entanglement witness refers to an operator $W$ that has non-negative expectation value $\text{Tr} \! \left( W \rho_\text{sep} \right) \geq 0$ for all separable states $\rho_\text{sep}$, and negative expectation value $\text{Tr} \! \left( W \sigma \right) < 0$ for at least one state $\sigma$. Any state that has a negative expectation value with respect to $W$ must therefore be entangled. An entanglement witness is referred to as optimal if there exists a separable state $\rho_\text{sep}$ such that $\text{Tr} \! \left( W \rho_\text{sep} \right) = 0$. 

We consider a system of $N$ qubits with a choice of two possible measurements on each particle,
\begin{align}
M^{(i)}_0 = \sigma^{(i)}_z , \quad M^{(i)}_1 = \sin \theta \, \sigma^{(i)}_x + \cos \theta \, \sigma^{(i)}_z .
\label{measurements}
\end{align}

\noindent Following the work by Tura~et~al.~\cite{tura14}, we construct an entanglement witness based on terms that are symmetric under all permutations of the particles. In particular, we define the symmetric two-body correlations $\mathcal{S}_{00}$, $\mathcal{S}_{01}$ and $\mathcal{S}_{11}$, where
\begin{align}
\mathcal{S}_{a b} = \sum\limits_{\mathclap{\substack{i,j=1 \\ i \neq j}}}^{N} M^{(i)}_a M^{(j)}_b. 
\label{symmcorrs}
\end{align}

\noindent The expectation values of these symmetric two-body correlations go to zero for the maximally mixed state $\mathbb{I} / 2^N$. 

These symmetric two-body correlations can then be used to construct an entanglement witness. We do so by showing (Appendix~\ref{app_construction}) that for all separable states, the expectation values satisfy the inequality
\begin{align}
\begin{split}
\frac{\alpha }{2} \left\langle \mathcal{S}_{00} \right\rangle + \beta \left\langle \mathcal{S}_{01} \right\rangle + \frac{\gamma}{2} \left\langle \mathcal{S}_{11} \right\rangle \leq F(\alpha, \beta, \gamma, \theta), \\
F(\alpha, \beta, \gamma, \theta) = \max \left\{ - N \lambda_z, (N^2 - N) \lambda_x \right\},
\end{split}
\label{sepboundeqn}
\end{align}

\noindent where
\begin{align}
\begin{split}
\lambda_z = \frac{1}{2} \left( A_{zz} + A_{xx} - \sqrt{\left( A_{zz} - A_{xx} \right)^2 + A_{zx}^2} \right), \\
\lambda_x = \frac{1}{2} \left( A_{zz} + A_{xx} + \sqrt{\left( A_{zz} - A_{xx} \right)^2 + A_{zx}^2} \right), \\
\end{split}
\label{lambdasdefn}
\end{align}

\noindent with
\begin{align}
\begin{split}
&A_{zz} = \frac{\alpha}{2} + \beta \cos \theta + \frac{\gamma}{2} \cos^2 \theta, \\
&A_{zx} = \beta \sin \theta + \frac{\gamma}{2} \sin 2 \theta , \\
&A_{xx} = \frac{\gamma}{2} \sin^2 \theta.
\end{split}
\label{coeffsdefn}
\end{align}

By defining the operator
\begin{align}
A(\theta) = \mathbb{I} - \frac{\frac{\alpha }{2} \mathcal{S}_{00} - \beta \mathcal{S}_{01} - \frac{\gamma}{2} \mathcal{S}_{11}}{F(\alpha, \beta, \gamma, \theta) },
\label{entwit}
\end{align}
we obtain an entanglement witness. The expectation value of this operator for all separable states is positive. For the even-$N$ case, the inequality in Eq.~(\ref{sepboundeqn}) is tight, meaning that it is saturated by at least one separable state (Appendix~\ref{app_construction}). Therefore, the entanglement witness $A(\theta)$ is optimal in the even-$N$ case, and there exists a separable state such that $\left\langle A(\theta) \right\rangle = 0$. On the other hand, when $N$ is odd, the inequality is not tight and thus $A(\theta)$ is not an optimal entanglement witness for odd $N$. As the value of $N$ increases, however, the entanglement witness becomes increasingly close to optimal for odd $N$.

\begin{figure*}
\subfloat[$N=4$, various values of $\theta$ \label{fig_entwitplots}]{
\includegraphics[width=0.23\textwidth, trim=.9cm 1.7cm 0.55cm 2.2cm, clip=true]{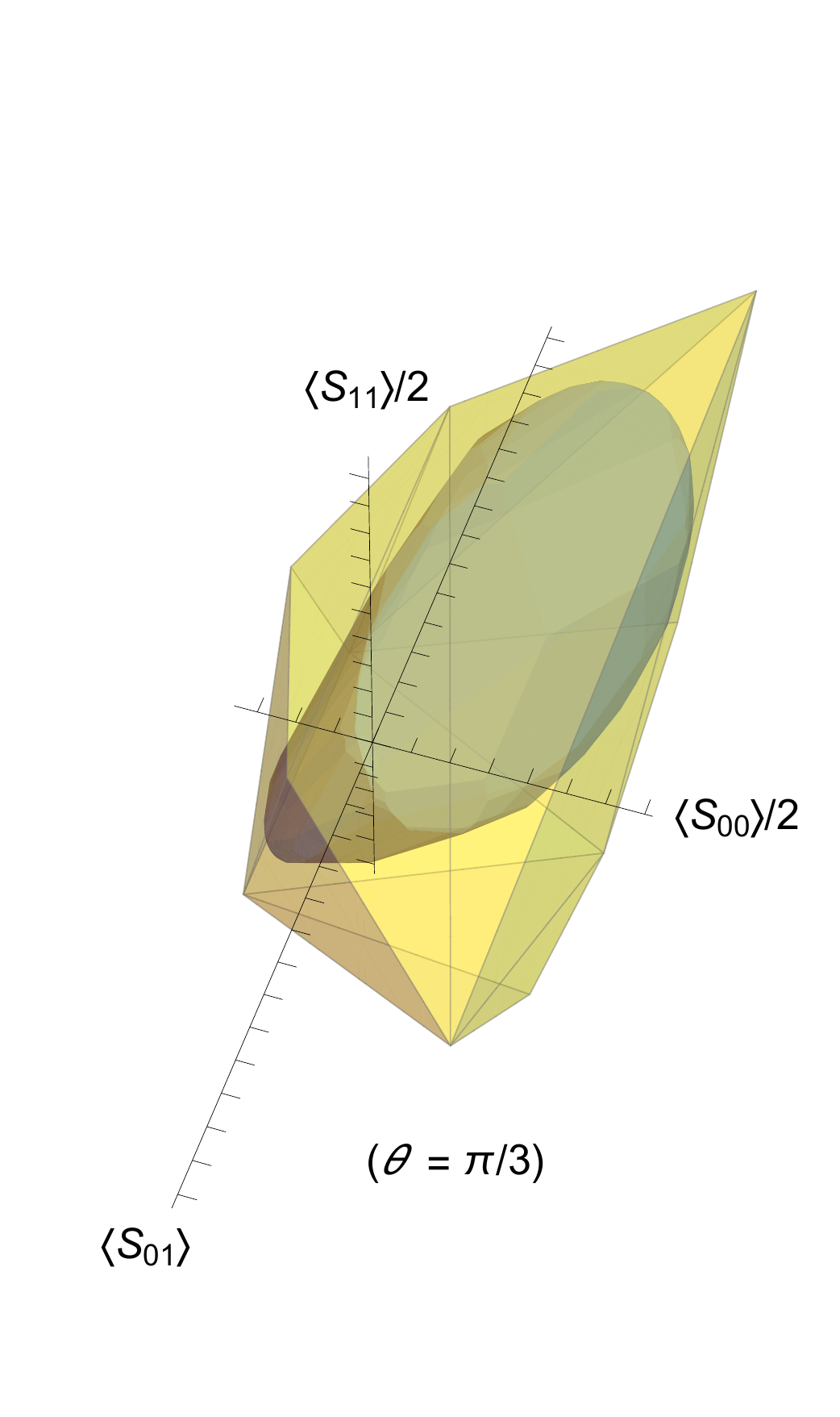}
\includegraphics[width=0.23\textwidth, trim=.9cm 1.7cm 0.55cm 2.2cm, clip=true]{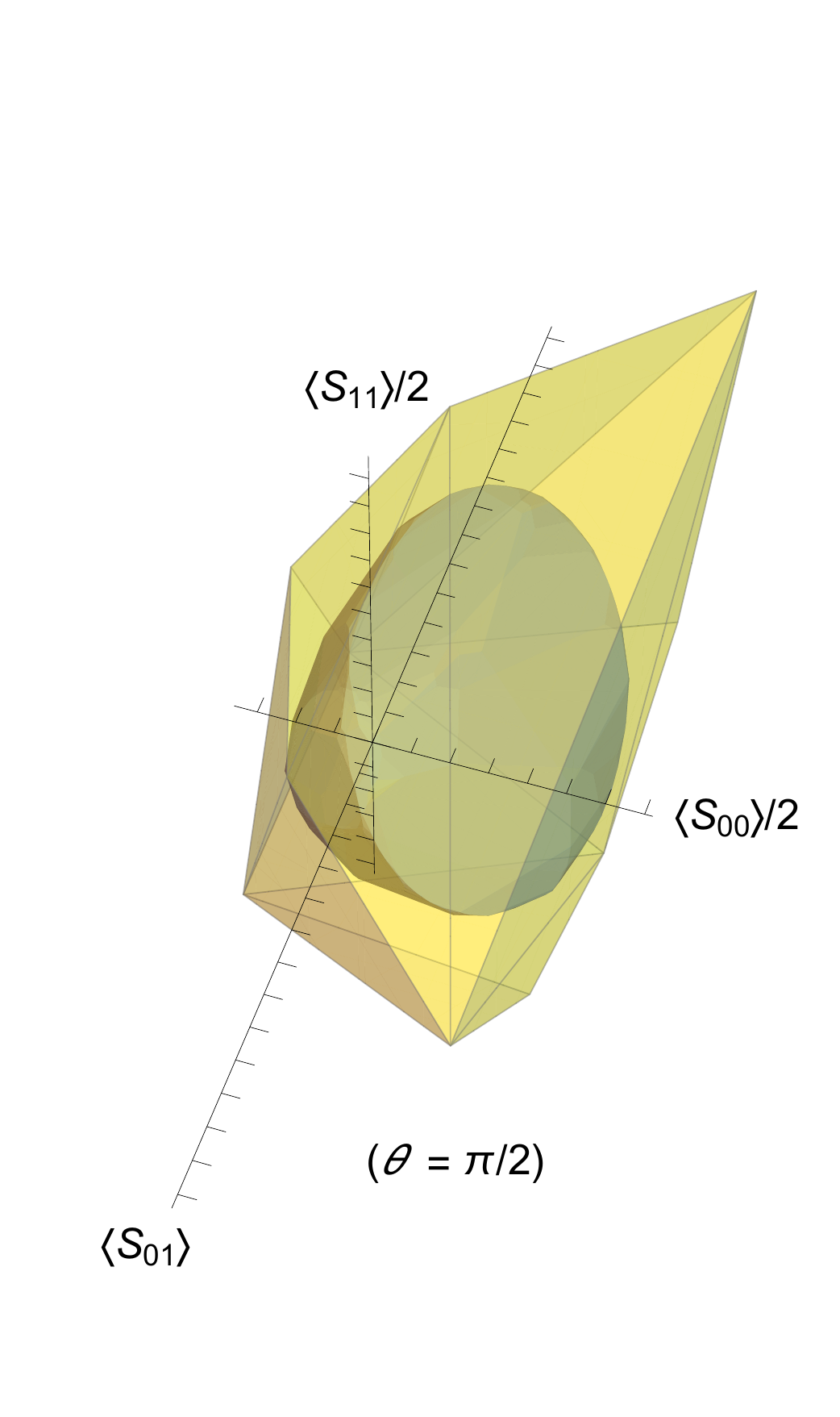}
\includegraphics[width=0.23\textwidth, trim=.9cm 1.7cm 0.55cm 2.2cm, clip=true]{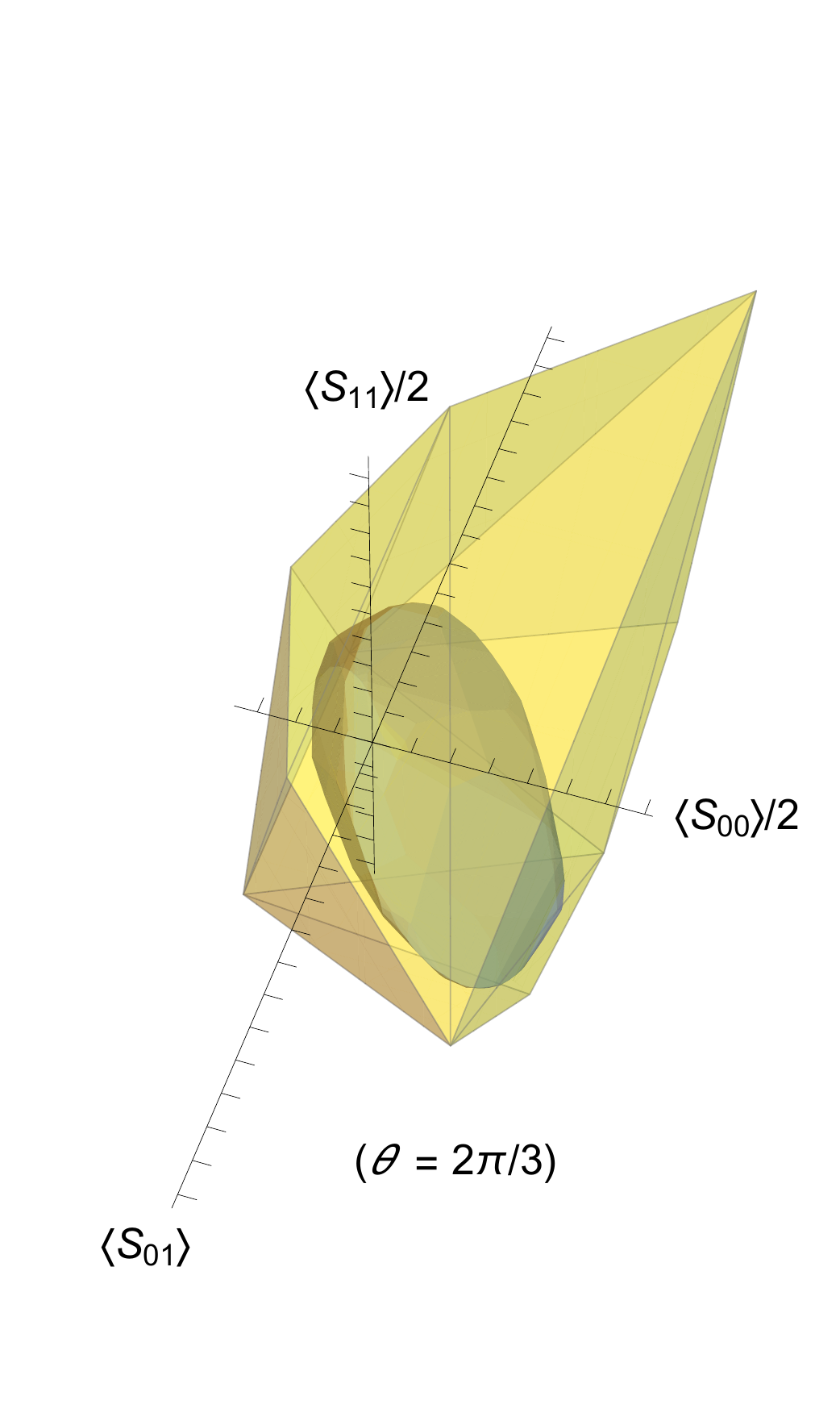}
}
\hspace{.4cm}
\subfloat[$N=3$, $\theta = \pi/3$ \label{fig_nonoptimal}]{
\includegraphics[width=0.24\textwidth, trim=0cm 3.3cm 0cm 2.3cm, clip=true]{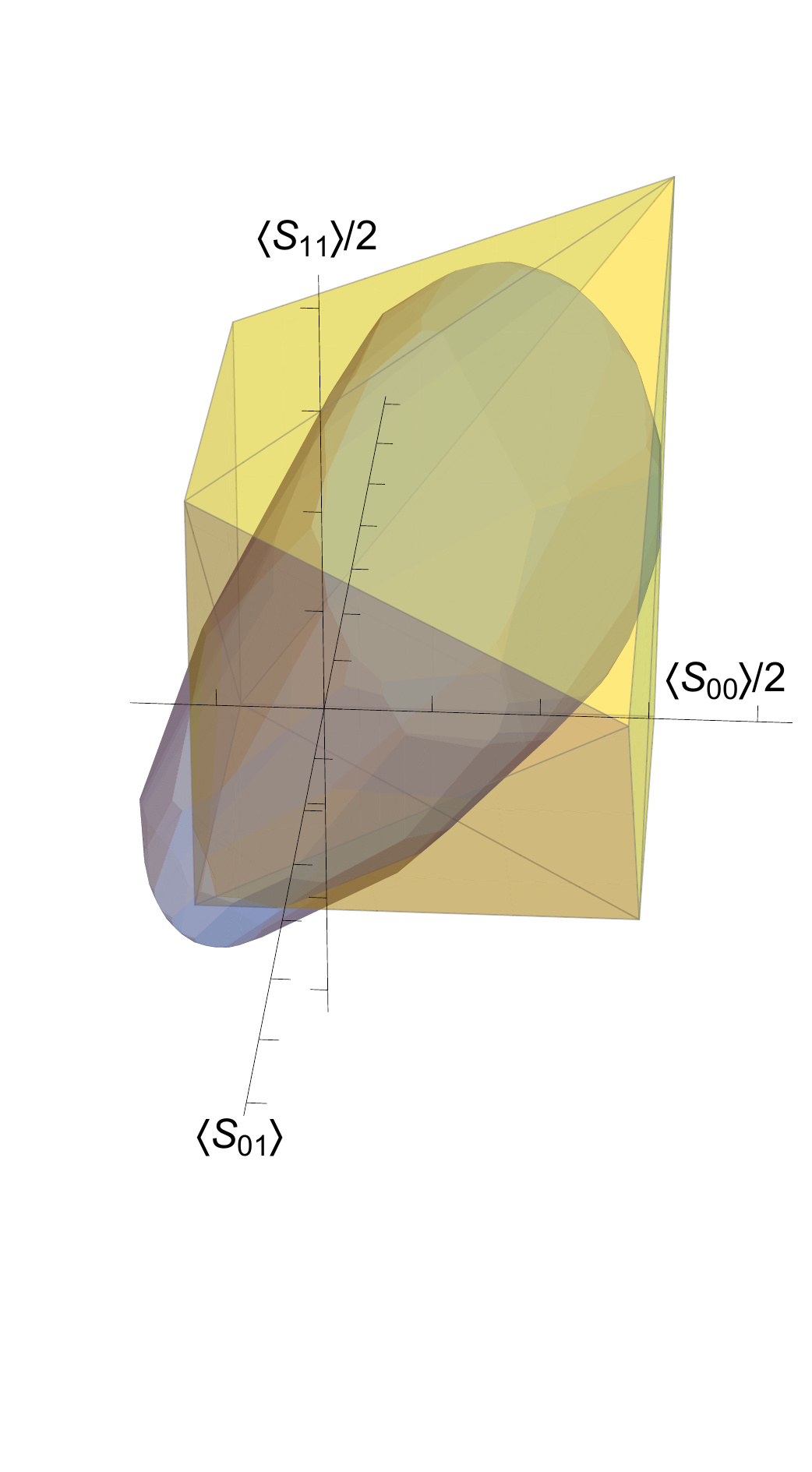}
}
\caption[]{(Colour online) Plots of the region in correlation space defined by the entanglement witness $A(\theta)$ for various values of $N$ and $\theta$. The blue region is an approximate depiction of the region defined by $A(\theta)$, while the yellow region shows the classical polytope. To reduce clutter, numerical values have been suppressed on the axes, and instead ticks have been placed at unit intervals. Since the value of $N$ is even in Fig.~\ref{fig_entwitplots}, $A(\theta)$ is an optimal entanglement witness for this case, and thus the blue region lies entirely within the classical polytope as expected. In contrast, Fig.~\ref{fig_nonoptimal} shows an odd value of $N$, for which the entanglement witness is non-optimal and the blue region exceeds the boundaries of the classical polytope.}
\end{figure*} 

Any scalar multiple of $A(\theta)$ is also an entanglement witness, though the normalisation in Eq.~(\ref{entwit}) is chosen such that $\left\langle A(\theta) \right\rangle = 1$ for the maximally mixed state. This choice of normalisation allows us to directly determine the robustness of entanglement detection against white noise, in that if some entangled state $\sigma$ has a negative expectation value $\left\langle A(\theta) \right\rangle = -Q$, then adding white noise to the state in the form
\begin{align}
\sigma_\text{noise} = (1-P) \sigma + P \, \frac{\mathbb{I}}{2^N}
\label{whitenoise}
\end{align}

\noindent causes the value of $\left\langle A(\theta) \right\rangle$ to become non-negative when $P \geq {Q}/(Q+1)$. Therefore, the larger the magnitude $Q$ of the negative expectation value, the greater the fraction $P$ of white noise that can be added before the state's remaining entanglement (if any) is no longer detected by the entanglement witness. 

\section{Geometric interpretation}\label{GI}

This entanglement witness can be interpreted geometrically by considering the symmetric two-body correlations $\left\{ \left\langle \mathcal{S}_{00} \right\rangle, \left\langle \mathcal{S}_{01} \right\rangle, \left\langle \mathcal{S}_{11} \right\rangle \right\}$ for any given state to correspond to a point in a 3-dimensional coordinate space, which may be referred to as a correlation space. For any specific choice of $\left\{ \alpha, \beta, \gamma \right\}$, the inequality in Eq.~(\ref{sepboundeqn}) then specifies a half-space such that all separable states lie within that half-space. The intersection of all such half-spaces forms a convex region in which all separable states are contained. If the entanglement witness is optimal, this region is precisely the region where all separable states lie, while if the witness is non-optimal, the latter is strictly contained in the former. 

This allows for a geometric interpretation of entanglement detection by $A(\theta)$. Given some quantum state, we can compute or experimentally obtain the values $\left\{ \left\langle \mathcal{S}_{00} \right\rangle, \left\langle \mathcal{S}_{01} \right\rangle, \left\langle \mathcal{S}_{11} \right\rangle \right\}$ and plot its position in correlation space, determining whether it lies within the region defined by $A(\theta)$. If it lies outside the region, it must be entangled, while if it lies inside the region, the result is inconclusive. Plotting the region defined by $A(\theta)$ encounters no scaling difficulties as $N$ increases, since evaluating $F(\alpha, \beta, \gamma, \theta)$ does not become more computationally intensive for larger values of $N$. 

A Bell inequality is an inequality that must be satisfied by any physical theory obeying the assumptions of locality and realism. Several examples of Bell inequalities based on symmetric one- and two-body correlations with two measurement settings were described and studied by Tura~et~al.~\cite{tura14}. Excluding the one-body correlations, these inequalities take the same form as Eq.~(\ref{sepboundeqn}), with various bounds on the right-hand side of the inequality. Similarly, each Bell inequality specifies a half-space in the correlation space, with the intersection of all such half-spaces forming a convex region which all local realistic models must lie within. It has been shown that the region defined in correlation space by all the Bell inequalities for a given system is a polytope~\cite{brunner14}; in other words, it is a convex set with only a finite number of extremal points. This polytope is variously known as the local polytope or classical polytope. Entangled states that lie outside the classical polytope are those that violate some Bell inequality.

Fig.~\ref{fig_entwitplots} shows plots of the region defined by $A(\theta)$ in comparison to the classical polytope in the $N=4$ case, for several values of $\theta$. In this as well as subsequent figures, the plotted regions are approximate representations in that only a finite number of tangent planes to the region were found and used to generate the plot. $N=4$ is an even value of $N$ and thus the entanglement witness is optimal for the plots shown in Fig.~\ref{fig_entwitplots}, with the depicted region specifying precisely the set of points corresponding to separable states. While no separable states can lie outside this region, there exist entangled states that lie within this region. Such entangled states cannot be detected by the entanglement witness $A(\theta)$, and a few examples of such states will be described later. It can be seen that the region defined by $A(\theta)$ lies entirely within the classical polytope, consistent with the fact that all separable states do not violate any Bell inequalities. 

In contrast, Fig.~\ref{fig_nonoptimal} shows an odd-$N$ case, specifically $N=3$. The non-optimality of the witness $A(\theta)$ can be seen from the fact that the blue region protrudes slightly from the classical polytope. Since the region where the separable states lie is supposed be contained within the classical polytope, this reflects the fact that the bound is non-optimal for this case, with the set of separable states being strictly contained within the blue region. For larger odd values of $N$, the extent to which the region defined by $A(\theta)$ exceeds the classical polytope decreases as the entanglement witness becomes more optimal.

\section{Dicke states and thermal states of LMG Hamiltonian}\label{Dicke}

An important class of entangled states that $A(\theta)$ can detect would be some of the Dicke states, 
\begin{align}
\ket{D^k_N} = s(\ket{N-k, k}),
\end{align}

\noindent where $\ket{i, j}$ denotes a pure product vector of $i$ qubits in the $\ket{0}$ state and $j$ qubits in the $\ket{1}$ state, while the function $s$ denotes symmetrisation over all particles along with appropriate normalisation. Specifically, Dicke states of the form $\ket{D^{\ceil{N/2}}_N}$ can have their entanglement detected by $A(\theta)$, and by extension the $\ket{D^{\floor{N/2}}_N}$ states can be detected as well, since they only differ from $\ket{D^{\floor{N/2}}_N}$ by local unitary rotations. Such Dicke states can arise as the ground states of the isotropic Lipkin-Meshkov-Glick (LMG) Hamiltonian~\cite{lipkin65}
\begin{align}
H_{LMG} = -\frac{\lambda}{N} \sum\limits_{\mathclap{\substack{i,j=1 \\ i < j}}}^{N} \left(\sigma^{(i)}_x \sigma^{(j)}_x +\sigma^{(i)}_y \sigma^{(j)}_y \right) + h \sum\limits_{i=1}^{N} \sigma^{(i)}_z,
\end{align}

\noindent which has $\ket{D^{\ceil{N/2}}_N}$ as its ground state~\cite{castanos06} when the parameters satisfy $\lambda/N \geq h > 0$. Some entanglement properties of the LMG Hamiltonian have been previously studied in terms of measures such as entanglement entropy~\cite{latorre05} and negativity~\cite{wichterich10}. Other techniques for the generation of Dicke states have also been proposed for systems such as Bose-Einstein condensates~\cite{raghavan01} and atomic populations~\cite{vanderbruggen11}, via heterodyne measurements.

\begin{figure}
\includegraphics[width=0.45\textwidth]{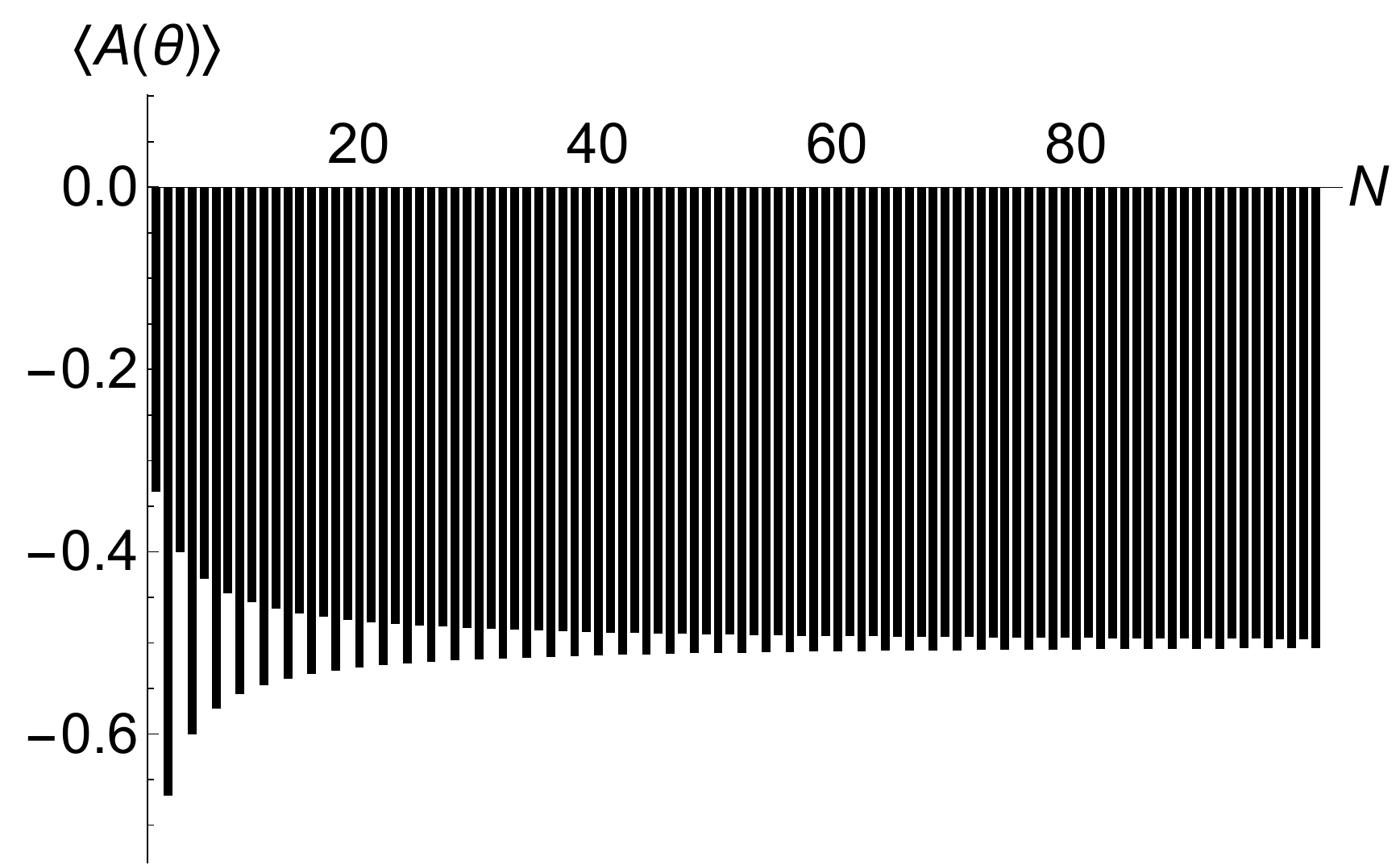}
\caption[]{Minimum values of $\left\langle A(\theta) \right\rangle$ with respect to the $\ket{D^{\ceil{N/2}}_{N}}$ Dicke states for $N=3$ to $N=100$. For even $N$, the magnitude of the values decreases as $N$ increases, while for odd $N$ the magnitude is increasing. This may reflect the increasing optimality of the entanglement witness $A(\theta)$ for large odd $N$. The values appear to converge towards $-0.5$ as $N$ becomes sufficiently large.}
\label{fig_entwitdicke}
\end{figure}

The expectation value of $A(\theta)$ with respect to the Dicke states can be computed as a closed-form expression (Appendix~\ref{app_matrix}). Fig.~\ref{fig_entwitdicke} shows the most negative values of $\left\langle A(\theta) \right\rangle$ for the $\ket{D^{\ceil{N/2}}_N}$ Dicke states as $N$ increases, minimised over $\alpha$, $\beta$, $\gamma$ and $\theta$. From the graph, it can be seen that for even values of $N$, the magnitude of the negative expectation value appears to be decreasing, while the opposite trend appears to hold for odd $N$. This latter likely reflects the fact that the entanglement witness is not optimal for odd $N$, but becomes increasingly optimal as $N$ increases. The graph also suggests that the values for odd and even $N$ converge towards $\left\langle A(\theta) \right\rangle \approx -0.5$ as $N$ increases, a proposal which is supported by numerical calculations of $\left\langle A(\theta) \right\rangle$ for larger values of $N$ than those shown in the figure. In terms of Eq.~\ref{whitenoise}, this allows $A(\theta)$ to tolerate a white noise fraction of $P = 1/3$ before it no longer detects the entanglement of $\ket{D^{\ceil{N/2}}_N}$. It may hence be suitable for use with large numbers of particles, since its entanglement detection is reasonably robust against white noise for all $N$.

Regarding the values of $\left\{ \alpha, \beta, \gamma, \theta \right\}$ which minimise $\left\langle A(\theta) \right\rangle$ for $\ket{D^{\ceil{N/2}}_N}$ states with large $N$, there appear to be multiple combinations of coefficients $\left\{ \alpha, \beta, \gamma \right\}$ which can yield $\left\langle A(\theta) \right\rangle \approx -0.5$, even after accounting for the fact that $A(\theta)$ remains unchanged if $\left\{ \alpha, \beta, \gamma \right\}$ are all multiplied by a positive constant. For any specific combination of $\left\{ \alpha, \beta, \gamma \right\}$, the range of values of $\theta$ for which $\left\langle A(\theta) \right\rangle $ remains negative decreases approximately with $1/\sqrt{N}$. For instance, the choice of coefficients $\alpha = 1$, $\beta = 1.13$, $\gamma = -1.14$ allows $\left\langle A(\theta) \right\rangle $ to be negative in the range $0 < \theta < 0.196$ when $N=100$, but only $0 < \theta < 0.0611$ when $N=1000$. This allows for some tolerance in misalignment of $\theta$ as long as $N$ is not too large.

\begin{figure*}
\subfloat[Minimum value of $\left\langle A(\theta) \right\rangle$ as a function of $T$ for the thermal state $\rho_T$. \label{fig_entwitthermal_graph}]{
\includegraphics[width=0.5\textwidth]{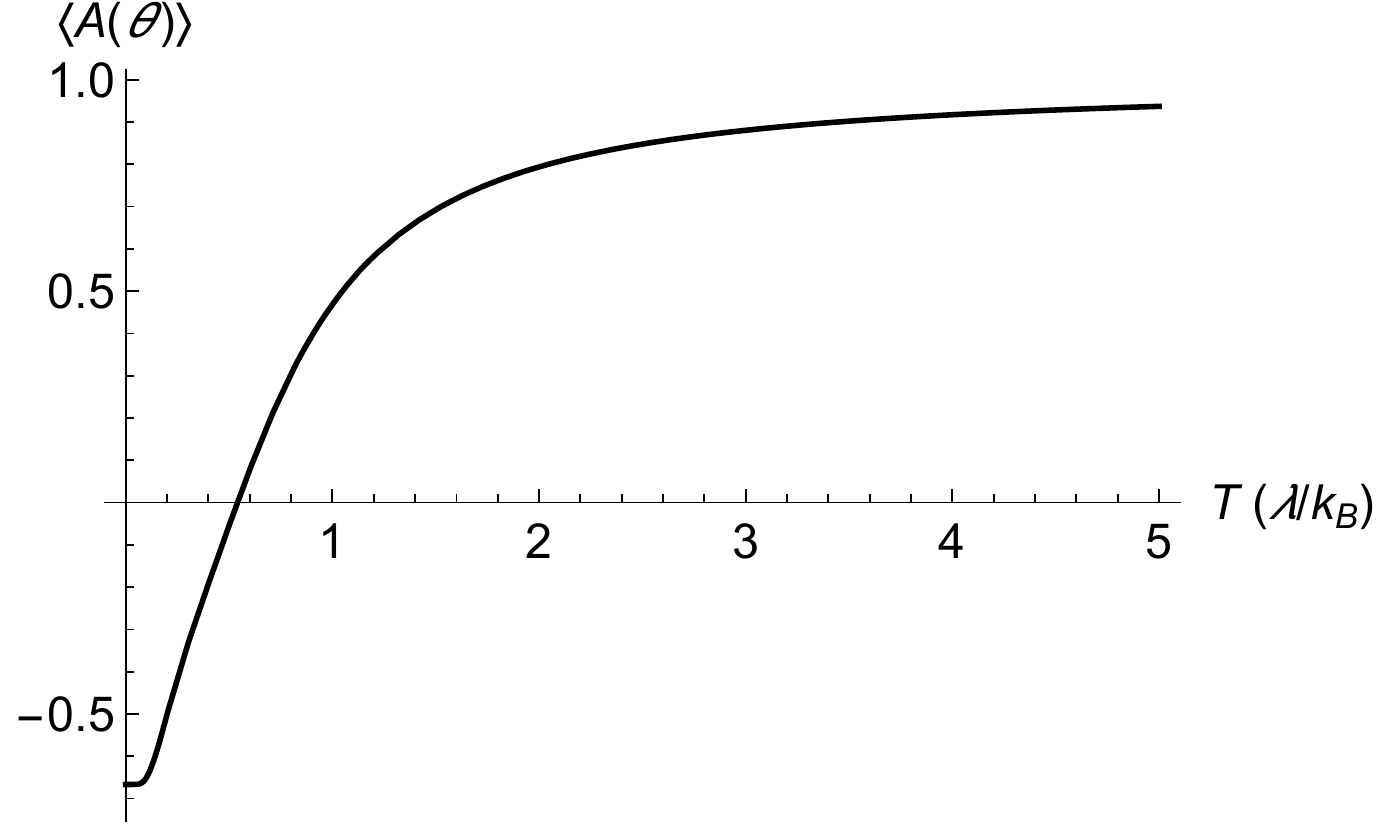}
}
\hspace{.5cm}
\subfloat[Coordinates in correlation space of the thermal state $\rho_T$. Numerical values have been suppressed on the axes, and instead ticks have been placed at unit intervals. \label{fig_entwitthermal_reg}]{
\includegraphics[width=0.4\textwidth, trim=0cm 3.5cm 0cm .8cm, clip=true]{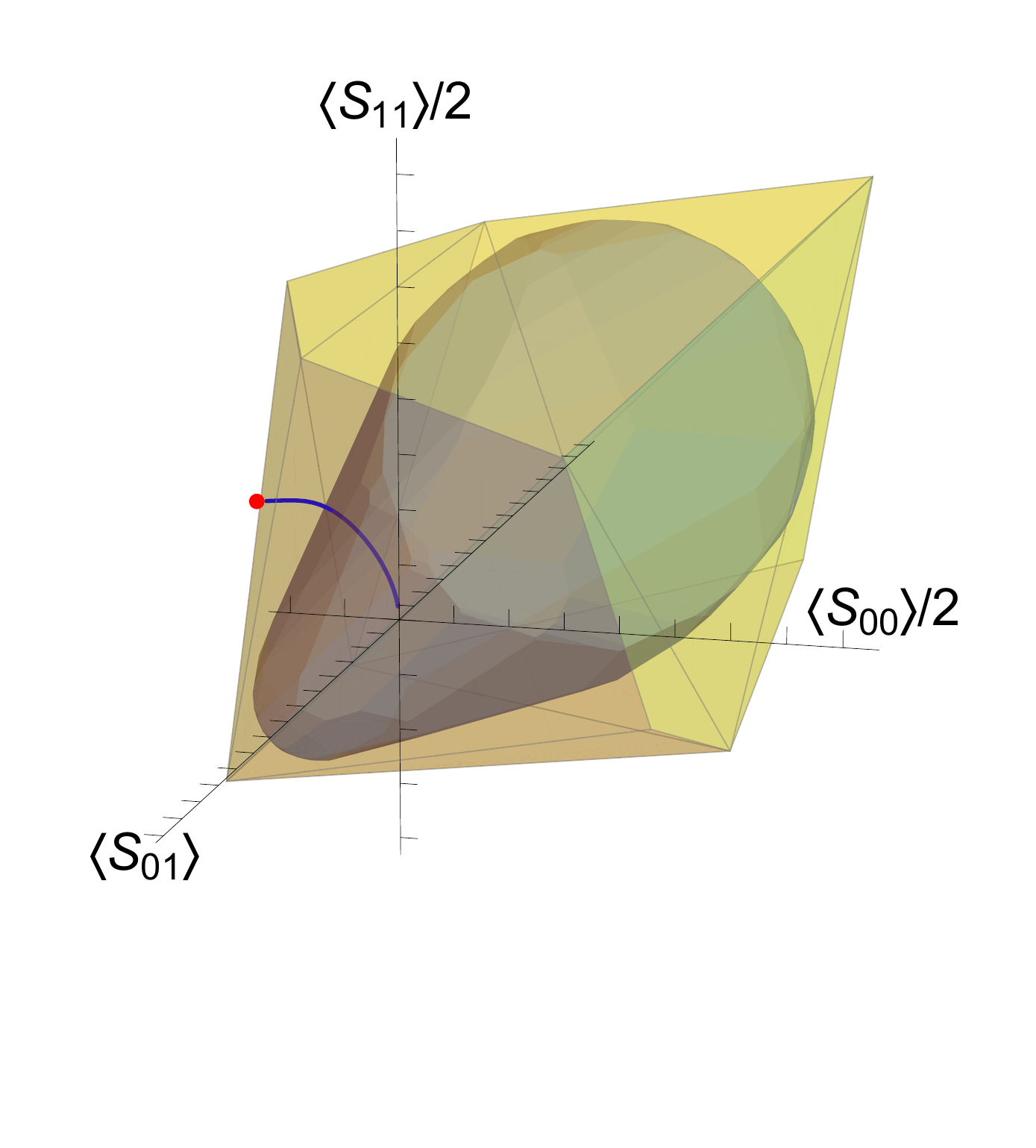}
}
\caption{(Colour online) Behaviour of the thermal state $\rho_T$ for $N=4$ with respect to $A(\theta)$. Fig.~\ref{fig_entwitthermal_graph} shows that the minimum value of $\left\langle A(\theta) \right\rangle$ is initially negative at $T=0$, becoming positive at a critical temperature of approximately $T_\text{crit} = 0.541 \lambda/k_B$. In Fig.~\ref{fig_entwitthermal_reg}, the region defined by $A(\theta)$ with $\theta =  \cos^{-1} (\ceil{N/2}/ (\ceil{N/2} + 1))$ is shown in blue, while the classical polytope is shown in yellow.  The red dot denotes the point corresponding to the thermal state $\rho_T$ at $T=0$, while the blue line leading towards the origin shows the locus of points corresponding to $\rho_T$ as $T$ increases.}
\label{fig_entwitthermal}
\end{figure*}

We also investigate the effect of thermal noise with respect to the LMG Hamiltonian, by considering the expectation value of $A(\theta)$ for the thermal state
\begin{align}
\rho_T = \frac{e^{- H_{LMG}/k_B T}}{\text{Tr} \! \left( e^{- H_{LMG}/k_B T} \right)}.
\end{align}

\noindent At $T=0$, this is the ground state $\ket{D^{\ceil{N/2}}_{N}}$. As $T$ goes to infinity, it approaches the maximally mixed state and thus becomes separable. 

Fig.~\ref{fig_entwitthermal_graph} shows the most negative value of $\left\langle A(\theta) \right\rangle$ as a function of $T$ for the thermal state $\rho_T$ with $N=4$ and $h=0.01 \lambda$. It can be seen from the graph that the minimum value of $\left\langle A(\theta) \right\rangle$ is indeed negative at $T=0$, consistent with the previously mentioned results for the $\ket{D^{\ceil{N/2}}_{N}}$ state. It becomes positive at a critical temperature of approximately $T_\text{crit} = 0.541 \lambda/k_B$, which can be taken as a measure of its robustness against such thermal noise.

Fig.~\ref{fig_entwitthermal_reg} shows the coordinates in correlation space of the thermal state $\rho_T$ with $N=4$. The value $\theta = \cos^{-1} (\ceil{N/2}/ (\ceil{N/2} + 1))$ was chosen to maximise the Bell inequality violation by $\ket{D^{\ceil{N/2}}_N}$, as shown by Tura~et~al.~\cite{tura14}. The ground state lies outside both the classical polytope and the region defined by $A(\theta)$, reflecting the fact that it violates a Bell inequality and can have its entanglement detected by $A(\theta)$. As the temperature increases, the thermal state moves towards the origin, entering the classical polytope followed by the region defined by $A(\theta)$ at the critical temperature $T_c$. While it lies within the former but not the latter, it is entangled and its entanglement can be detected by $A(\theta)$, but it does not violate any Bell inequalities based on symmetric two-body correlations. 

Apart from the $\ket{D^k_N}$ states with $k = \ceil{N/2}$ or $k = \floor{N/2}$, the entanglement witness $A(\theta)$ is also able to detect the entanglement of Dicke states with other values of $k$ sufficiently close to $N/2$. The smallest value of $N$ for which this occurs is $N=18$, where the states $\ket{D^8_{18}}$ and $\ket{D^{10}_{18}}$ have minimum expectation value $\left\langle A(\theta) \right\rangle = -0.301$. This result can be improved upon by generalising the measurement settings in Eq.~\ref{measurements} to allow both measurements to be in arbitrary directions (Appendix~\ref{app_generalisation}), in which case the smallest value of $N$ for which this occurs is $N=8$, where $\ket{D^3_{8}}$ and $\ket{D^{5}_{8}}$ yield minimum expectation value $\left\langle A(\theta) \right\rangle = -0.0714$. With this generalised version, the $\ket{D^7_{18}}$ and $\ket{D^{11}_{18}}$ states are detected as well. These trends are similar to those for the spin-squeezing inequalities described in~\cite{toth09}, which are maximally violated by $\ket{D^{N/2}_N}$ states, though the inequalities in that case are also able to detect any entangled Dicke state.

\section{Maximal entanglement detection and spin-squeezed states}\label{Maximal}

The Dicke states themselves are not the states which yield the lowest possible value of $\left\langle A(\theta) \right\rangle$. However, it can be shown that the minimum value of $\left\langle A(\theta) \right\rangle$ is always achieved by a symmetric state, and the Dicke states form a basis for the symmetric subspace. The matrix elements of $A(\theta)$ with respect to this subspace can be explicitly derived (Appendix~\ref{app_matrix}), and thus the lowest possible value of $\left\langle A(\theta) \right\rangle$ can be found as the minimum eigenvalue of this matrix, with the state that achieves this value being a symmetric eigenstate of $A(\theta)$. Numerical data suggests as $N$ increases, the minimum value of $\left\langle A(\theta) \right\rangle$ decreases asymptotically towards a limiting value of $\left\langle A(\theta) \right\rangle = -1$. This corresponds to $A(\theta)$ being able to tolerate a white noise fraction of $P=1/2$ before failing to detect the entanglement of this eigenstate. We note that since this eigenstate is symmetric, it is entangled if and only if it is genuinely multipartite entangled~\cite{ichikawa08}. Similar results hold for the entanglement witness generalised to measurements along arbitrary directions (Appendix~\ref{app_generalisation}).

We have found that this maximal detection can be approached by spin-squeezed states~\cite{kitagawa93},
\begin{align}
\ket{S_N(\chi)} = \frac{1}{\sqrt{2^N}} \sum\limits_{k=0}^{N} \sqrt{\begin{pmatrix} N \\ k \end{pmatrix}} e^{-i \chi (k-N/2)^2} \ket{D^k_N}.
\label{spinsq}
\end{align}

\noindent When $\chi=0$, this is simply the state with all spins aligned along the $x$-direction. Spin-squeezed states are of significance because for appropriate values of $\chi$, the variance of the spin component along one axis can be reduced, at the expense of increasing the variance of the spin component along an orthogonal axis. They have also given rise to spin-squeezing inequalities for witnessing entanglement~\cite{sorenson01,korbicz06,toth07,toth09,lucke14}. A number of methods have been studied for the generation of spin-squeezed states~\cite{riedel10,gross10,vanderbruggen11}.

While $A(\theta)$ as defined in Eq.~(\ref{entwit}) does not detect the entanglement of spin-squeezed states of the form in Eq.~(\ref{spinsq}), the generalised version for arbitrary measurement directions (Appendix~\ref{app_generalisation}) detects such entanglement strongly, particularly for large values of $N$. The minimum expectation value is $-0.430$ for $N=3$, and reaches $-0.940$ when $N=1000$. This appears to be approaching the aforementioned bound of $-1$, suggesting that spin-squeezed states come close to being maximally detected by this entanglement witness. This may facilitate experimental investigations of this entanglement witness, given that various schemes have been developed to generate spin-squeezed states. We note, however, that the value of $\chi$ for which the expectation value is minimised decreases asymptotically towards zero as $N$ increases, but the spin-squeezed state with precisely $\chi = 0$ is always separable. This suggests that the range of values of $\chi$ for which $\ket{S_N(\chi)}$ is detected becomes narrower as $N$ increases.

\section{Detection of superposition states}\label{Superposed}

There exist some classes of entangled states which cannot be detected by $A(\theta)$, such as the Greenberger-Horne-Zeilinger (GHZ) states, 
\begin{align}
\ket{GHZ_N} = \frac{1}{\sqrt{2}} \left( \ket{D^0_N} + \ket{D^N_N} \right),
\end{align}

\noindent for any $N \geq 3$. More generally, the entanglement of an $N$-particle GHZ state cannot be detected using any $k$-body correlations with $k<N$, because the set of states obtained by tracing out any non-zero number of particles from the GHZ state is indistinguishable from the set of states obtained by tracing out particles from the separable state $\rho = \left( \ket{D^0_N} \bra{D^0_N} + \ket{D^N_N} \bra{D^N_N} \right) / 2$. For similar reasons, another example of entanglement that cannot be detected using two-body correlations would be the four-particle Smolin state~\cite{smolin01},
\begin{align}
\rho_S = \frac{1}{4} \sum\limits_{\mu = 1}^{4} \ket{\Psi^{(1,2)}_\mu} \bra{\Psi^{(1,2)}_\mu} \otimes \ket{\Psi^{(3,4)}_\mu} \bra{\Psi^{(3,4)}_\mu},
\label{smolin}
\end{align}

\noindent where $\ket{\Psi^{(i,j)}_\mu}$ with $\mu = 1,2,3,4$ refer to the four Bell states with respect to particles $i$ and $j$. The values of $\left\langle \mathcal{S}_{00} \right\rangle$, $\left\langle \mathcal{S}_{01} \right\rangle$ and $\left\langle \mathcal{S}_{11} \right\rangle$ are in fact zero for the Smolin state, and thus it lies at the origin of the correlation space, which is always within the region defined by $A(\theta)$.

However, we have found that $A(\theta)$ is able to detect the superposition state $\cos\Omega \ket{D^2_N} + \sin\Omega \ket{GHZ_N}$ over a significant range of values of $\Omega$, despite the fact that it cannot detect the GHZ states or $\ket{D^2_N}$ states by themselves for $N > 5$. The minimum value of $\left\langle A(\theta) \right\rangle$ with respect to this state appears to approach $-0.235$ for large values of $N$, corresponding to a reasonable degree of robustness against white noise. The range of values of $\Omega$ for which $\left\langle A(\theta) \right\rangle$ remains negative approaches $\pi/2 < \Omega < 2.04$ when $N$ is large, with the most negative value occurring at approximately $\Omega = 1.80$. Since $\Omega = \pi/2$ corresponds to the GHZ state itself, this indicates that superposing the GHZ state with a small $\ket{D^2_N}$ component suffices to allow $A(\theta)$ to detect its entanglement. In terms of the correlation space, this arises because for any $N$, the GHZ state lies on the boundary of the region defined by $A(\theta)$, and superposing it with the $\ket{D^2_N}$ state causes it to trace out a curve that leaves the region defined by $A(\theta)$, allowing its entanglement to be detected. 

\section{Conclusion}\label{Conclusion}

To conclude, we have developed an entanglement witness based on symmetric two-body correlations, which are more experimentally accessible as compared to higher-order correlations. The entanglement witness admits a geometric interpretation which encounters no scaling difficulties as the number of particles increases, and is able to detect the entanglement of some Dicke states with an expectation value that approaches a constant value of $\left\langle A(\theta) \right\rangle \approx -0.5$ for large $N$. Such Dicke states can be realised as ground states of the LMG Hamiltonian or via other techniques~\cite{raghavan01,vanderbruggen11}, and we have shown that the entanglement detection is robust against some thermal noise. This may allow for experimental investigation of these results. The entangled states that are most strongly detected have an expectation value that approaches $\left\langle A(\theta) \right\rangle \approx -1$ for large $N$, corresponding to a reasonable degree of robustness against white noise. Spin-squeezed states are able to approach this maximal detection, which may provide another avenue for experimental implementation~\cite{riedel10,gross10,vanderbruggen11}. The entanglement witness is also able to detect states other than Dicke states; in particular, we showed that the witness could detect states that are in superposition of Dicke states with GHZ states, even though the latter states are not detectable with the witness. 

The authors acknowledge useful correspondence with Antonio Acin and Valerio Scarani. This project is supported by the National Research Foundation, Singapore, and the Ministry of Education, Singapore.

\bibliographystyle{apsrev4-1} 
\bibliography{twobodywitness_biblio} 

\clearpage

\appendix

\section{Construction of entanglement witness}
\label{app_construction}

To prove Eq.~(\ref{sepboundeqn}), we begin by noting that it suffices to find the maximum of $\frac{\alpha }{2} \left\langle \mathcal{S}_{00} \right\rangle + \beta \left\langle \mathcal{S}_{01} \right\rangle + \frac{\gamma}{2} \left\langle \mathcal{S}_{11} \right\rangle$ for all pure separable states, rather than having to consider mixed separable states as well. This is because any separable mixed state $\rho_\text{sep}$ can be written as a convex combination of product states,
\begin{align}
&\rho_\text{sep} = \sum\limits_{i=1}^{k} p_i \, \rho_{i}^{(1)} \otimes \rho_{i}^{(2)} \otimes ... \otimes \rho_{i}^{(N)}, \\
&\text{with } p_i > 0, \, \sum\limits_{i=1}^{M} p_i = 1, \nonumber
\label{sepdefn}
\end{align}

\noindent such that all the individual qubit states $\rho_i^{(j)}$ are pure. It can then be seen that if the inequality in Eq.~(\ref{sepboundeqn}) is satisfied for all pure separable states, then any mixed separable state $\rho_\text{sep}$ also satisfies the inequality.

For a pure separable state $\rho^{(1)} \otimes \rho^{(2)} \otimes ... \otimes \rho^{(N)} $, the state $\rho^{(i)}$ of each qubit is characterised by a Bloch vector $\hat{n}_i$ of norm 1,
\begin{align}
\rho^{(i)} = \frac{\mathbb{I} + \hat{n}_i \cdot \vec{\sigma}}{2}, 
\end{align}

\noindent where $\vec{\sigma}$ is the Pauli vector $(\sigma_x, \sigma_y, \sigma_z)$. Using the fact that $\left\langle \vec{\sigma} \right\rangle_{\rho^{(i)}} = \hat{n}_i$, the expectation values of the symmetric two-body correlations can be expressed in terms of the Bloch vector components $\hat{n}_i = \left(x_i, y_i, z_i \right)$: 
\begin{align}
& \left\langle \mathcal{S}_{00} \right\rangle = \sum\limits_{\mathclap{\substack{i,j=1 \\ i \neq j}}}^{N} z_i z_j, \\
& \left\langle \mathcal{S}_{01} \right\rangle = \sum\limits_{\mathclap{\substack{i,j=1 \\ i \neq j}}}^{N} \left( \sin \theta \, z_i x_j + \cos \theta \, z_i z_j \right), \\
& \left\langle \mathcal{S}_{11} \right\rangle = \sum\limits_{\mathclap{\substack{i,j=1 \\ i \neq j}}}^{N} \left( \sin^2 \theta \, x_i x_j + \cos^2 \theta \, z_i z_j + \sin 2 \theta \, z_i x_j \right),
\end{align}

\noindent using the measurement settings given in Eq.~(\ref{measurements}).

In addition, we note that 
\begin{align}
\sum\limits_{\mathclap{\substack{i,j=1 \\ i \neq j}}}^{N} z_i z_j = N^2 \overline{z}^2 - N \overline{z^2},
\end{align}

\noindent where we introduce the notation $\overline{f(z)} = \frac{1}{N} \sum_{i=1}^{N} f(z_i)$ for any function $f$. Essentially, this is an average over the Bloch components of the individual qubits. A similar result holds for $\overline{x}^2$ with $\overline{x^2}$, as well as $\overline{z} \,\, \overline{x}$ with $\overline{z x}$. Combining these results, we find that
\begin{widetext}
\begin{align}
\left\langle \frac{\alpha }{2} \mathcal{S}_{00} + \beta \mathcal{S}_{01} + \frac{\gamma}{2} \mathcal{S}_{11} \right\rangle &= \sum\limits_{\mathclap{\substack{i,j=1 \\ i \neq j}}}^{N} \left( A_{zz} z_i z_j + A_{zx} z_i x_j + A_{xx} x_i x_j \right) \nonumber \\
&= N^2 \left( A_{zz} \overline{z}^2 + A_{zx} \overline{z} \, \overline{x} + A_{xx} \overline{x}^2 \right) - N \left( A_{zz} \overline{z^2} + A_{zx} \overline{z x} + A_{xx} \overline{x^2} \right), 
\end{align}
\end{widetext}

\noindent where $A_{zz}, A_{zx}, A_{xx}$ have been introduced as in Eq.~(\ref{coeffsdefn}). 

We now make use of the fact that any quadratic form can be diagonalised with an orthogonal transformation. Specifically, there exists a $2 \times 2$ orthogonal matrix $M$, depending only on $A_{zz}$, $A_{zx}$ and $A_{xx}$, such that for all $z, x$ we have
\begin{align}
A_{zz} z^2 + A_{zx} z x + A_{xx} x^2 = \lambda_z z'^2 + \lambda_x x'^2,
\end{align}

\noindent where $(z' \; x')^T = M^{-1} (z \; x)^T$ and $\lambda_z$, $\lambda_x$ are defined as in Eq.~(\ref{lambdasdefn}). $\lambda_z$, $\lambda_x$ are essentially the eigenvalues associated with the quadratic form. By applying the change of coordinates $(z_i \; x_i)^T = M (z'_i \; x'_i)^T$ to the individual Bloch vectors, it can hence be shown that
\begin{align}
A_{zz} \overline{z^2} + A_{zx} \overline{z x} + A_{xx} \overline{x^2} = \lambda_z \overline{z'^2} + \lambda_x \overline{x'^2}, \\
A_{zz} \overline{z}^2 + A_{zx} \overline{z} \, \overline{x} + A_{xx} \overline{x}^2 = \lambda_z \overline{z'}^2 + \lambda_x \overline{x'}^2.
\end{align}

We hence wish to maximise 
\begin{align}
N^2 \left( \lambda_z \overline{z'}^2 + \lambda_x \overline{x'}^2 \right) - N \left( \lambda_z \overline{z'^2} + \lambda_x \overline{x'^2} \right),
\label{diagform}
\end{align}

\noindent subject to the constraints $z_i^2 + x_i^2 \leq 1$. Because the transformation was orthogonal, we have $z'^2_i + x'^2_i = z_i^2 + x_i^2$ and thus the constraints can be equivalently stated as $z'^2_i + x'^2_i \leq 1$. Noting that $\overline{z'^2} - \overline{z'}^2$ is essentially the variance of the set $\left\{ z'_i \right\}$ and thus $\overline{z'^2} - \overline{z'}^2 \geq 0$ with equality if and only if all $z'_i$ have the same value, we see that
\begin{align}
\begin{split}
N^2 \overline{z'}^2 - N \overline{z'^2} & \leq (N^2 - N) \overline{z'^2}, \\
N^2 \overline{z'}^2 - N \overline{z'^2} & \geq -N \overline{z'^2}.
\end{split}
\end{align}

\noindent The upper bound is achieved if and only if all $z'_i$ have the same value, while the lower bound is achieved if and only if $\overline{z'} = 0$. These bounds can be summarised as stating that for any $\lambda_z$, we have
\begin{align}
\lambda_z \left( N^2 \overline{z'}^2 - N \overline{z'^2} \right) \leq \max \left\{ (N^2 - N) \lambda_z , -N \lambda_z \right\} \overline{z'^2}. 
\end{align}

\noindent A similar statement holds for the $x'$ terms, leading to the final bound on Eq.~(\ref{diagform}),
\begin{widetext}
\begin{align}
\left\langle \frac{\alpha }{2} \mathcal{S}_{00} + \beta \mathcal{S}_{01} + \frac{\gamma}{2} \mathcal{S}_{11} \right\rangle &= \lambda_z \left( N^2 \overline{z'}^2 - N \overline{z'^2} \right) + \lambda_x \left( N^2 \overline{x'}^2 - N \overline{x'^2} \right) \nonumber \\
& \leq \max \left\{ (N^2 - N) \lambda_z, - N \lambda_z \right\} \overline{z'^2} + \max \left\{ (N^2 - N) \lambda_x , - N \lambda_x \right\} \overline{x'^2} \nonumber \\
& \leq \max \left\{ (N^2 - N) \lambda_z , - N \lambda_z, (N^2 - N) \lambda_x , - N \lambda_x \right\} \nonumber \\
&= \max \left\{ - N \lambda_z, (N^2 - N) \lambda_x \right\} \quad \text{since $\lambda_z \leq \lambda_x$.}
\label{sepbound}
\end{align}
\end{widetext}

\noindent The penultimate step is achieved by noting that $C_z z'^2_i + C_x x'^2_i \leq \max \left\{ C_z, C_x \right\}$ under the constraint $z'^2_i + x'^2_i \leq 1$, which can be proven via Lagrange multipliers or by viewing it as essentially finding the extremal points of an ellipse. 

We have thus found that the function $F(\alpha, \beta, \gamma, \theta) =\max \left\{ - N \lambda_z, (N^2 - N) \lambda_x \right\}$ satisfies the condition in Eq.~(\ref{sepboundeqn}), and can hence be used to construct an entanglement witness $A(\theta)$ as in Eq.~(\ref{entwit}). When $N$ is even, this bound is tight, as we now show by explicitly constructing a separable state that saturates this inequality. For the case where the maximum of the quantities in Eq.~(\ref{sepbound}) is $ (N^2 - N) \lambda_x $, the bound is achieved by the state with $z'_i = 0$, $x'_i = 1$ for all the qubits. This can then be converted back into values for the original Bloch components $z_i$ and $x_i$ by inverting the orthogonal transformation $(z_i \; x_i)^T = M (z'_i \; x'_i)^T$.  For the case where the maximum is $- N \lambda_z $, the bound is achieved by the state with $z'_i = 1$, $x'_i = 0$ for half the qubits and $z'_i = -1$, $x'_i = 0$ for the remaining half, yielding $\overline{z'^2} = 1$ and $\overline{z'} = 0$. 

If $N$ is odd, the conditions $\overline{z'^2} = 1$ and $\overline{z'} = 0$ cannot be fulfilled simultaneously, and thus the inequality cannot be saturated for certain combinations of values for $\lambda_z$, $\lambda_x$. However, we note that for increasing values of odd $N$, the bound becomes increasingly tight, because when $N$ is large it is possible to approach $\overline{z'^2} \approx 1$ and $\overline{z'} \approx 0$ even when $N$ is odd. This entanglement witness thus becomes closer to optimal for odd $N$ when $N$ is large.\\

\section{Matrix elements of $A(\theta)$ in symmetric subspace}
\label{app_matrix}

The Dicke states form a basis for the symmetric subspace. By considering $\sigma_x$ to act as a bit-flip operation, we can compute the matrix elements of $A(\theta)$ in this subspace. For any $i \neq j$, $k \geq k'$, the only nonzero matrix elements are
\begin{widetext}
\begin{align}
\begin{split}
& \bra{D^k_N} \sigma^{(i)}_z \sigma^{(j)}_z \ket{D^{k'}_N} = \tfrac{N^2-N-4kN+4k^2}{N^2-N} \; \text{ for } k - k' = 0,\\
& \bra{D^k_N} \sigma^{(i)}_z \sigma^{(j)}_x \ket{D^{k'}_N} = \tfrac{(N-2k+1) \sqrt{k(N-k+1)}}{N^2-N} \; \text{ for } k - k' = 1,\\
& \bra{D^k_N} \sigma^{(i)}_x \sigma^{(j)}_x \ket{D^{k'}_N} = \left\{ 
\begin{array}{lr}
\frac{\sqrt{k(k-1)(N-k+2)(N-k+1)}}{N^2-N} &\text{for } k - k' = 2,\\
\frac{2k(N-k)}{N^2-N} &\text{for } k - k' = 0.\\
\end{array}
\right.
\end{split}
\end{align}
\end{widetext}

\noindent The matrix elements for $k < k'$ can also be directly seen from these formulas since the matrix is Hermitian. The matrix elements of the symmetric two-body correlations $\left\{ \left\langle \mathcal{S}_{00} \right\rangle, \left\langle \mathcal{S}_{01} \right\rangle, \left\langle \mathcal{S}_{11} \right\rangle \right\}$ in this subspace are then easily derived by noting that all $(N^2-N)$ terms in the summations in Eq.~(\ref{symmcorrs}) have the same matrix elements in this subspace, since the above formulas apply for all $i \neq j$. It can be seen that the matrix representation of $A(\theta)$ is pentadiagonal and has dimensions $(N+1) \times (N+1)$, which makes it possible to evaluate its eigenvalues numerically. The expectation value of $A(\theta)$ with respect to any Dicke state can be obtained by taking the diagonal elements.

\section{Generalisation to arbitrary measurement directions}
\label{app_generalisation}

We consider measurement settings along arbitrary directions,
\begin{align}
\begin{split}
M^{(i)}_0 = \sin \theta_0 \cos \phi_0 \, \sigma^{(i)}_x + \sin \theta_0 \sin \phi_0 \, \sigma^{(i)}_y + \cos \theta_0 \, \sigma^{(i)}_z , \\
M^{(i)}_1 = \sin \theta_1 \cos \phi_1 \, \sigma^{(i)}_x + \sin \theta_1 \sin \phi_1 \, \sigma^{(i)}_y + \cos \theta_1 \, \sigma^{(i)}_z .
\end{split}
\label{measurements2}
\end{align}

\noindent Following a similar argument as that used in Appendix~\ref{app_construction}, we can then show that
\begin{multline}
\frac{\alpha }{2} \left\langle \mathcal{S}_{00} \right\rangle + \beta \left\langle \mathcal{S}_{01} \right\rangle + \frac{\gamma}{2} \left\langle \mathcal{S}_{11} \right\rangle \leq \\
\max \left\{ (N^2 - N) \lambda_i, - N \lambda_i \right\}_{i \in \{ x, y, z\}},
\end{multline}

\noindent with $\left\{ \lambda_z, \lambda_y, \lambda_x \right\}$ being the eigenvalues of a $3 \times 3$ symmetric matrix $B$, which can be computed as 
\begin{align}
B = \frac{\alpha }{2} m_0^T m_0 + \frac{\beta}{2} \left( m_0^T m_1 + m_1^T m_0 \right) + \frac{\gamma}{2} m_1^T m_1 ,
\end{align}

\noindent introducing the row vectors
\begin{align}
\begin{split}
m_0 = \left( \sin \theta_0 \cos \phi_0 , \sin \theta_0 \sin \phi_0 , \cos \theta_0 \right), \\
m_1 = \left( \sin \theta_1 \cos \phi_1 , \sin \theta_1 \sin \phi_1 , \cos \theta_1 \right). 
\end{split}
\end{align}

The line of reasoning is essentially identical to that used in Appendix~\ref{app_construction}, except that the quadratic form is in three variables $\left\{ x, y, z \right\}$ rather than two. Computing $\left\{ \left\langle \mathcal{S}_{00} \right\rangle, \left\langle \mathcal{S}_{01} \right\rangle, \left\langle \mathcal{S}_{11} \right\rangle \right\}$ with these measurement settings in the symmetric subspace can be simplified using the matrix elements $\bra{D^k_N} \sigma^{(i)}_x \sigma^{(j)}_y \ket{D^{k'}_N}$, $\bra{D^k_N} \sigma^{(i)}_y \sigma^{(j)}_y \ket{D^{k'}_N}$, $\bra{D^k_N} \sigma^{(i)}_z \sigma^{(j)}_y \ket{D^{k'}_N}$, which can be evaluated by means similar to those in Appendix~\ref{app_matrix}.

\end{document}